# Noninvasive linear fluorescence imaging through scattering media via wavefront shaping.

Anat Daniel[1,†], Dan Oron[1], Yaron Silberberg[1]

*Affiliation:* [1] Department of Physics of Complex Systems, Weizmann Institute of Science, Rehovot, Israel

**We demonstrate focusing and imaging through a scattering medium noninvasively by using wavefront shaping. Our concept is based on utilizing the spatial fluorescence contrast which naturally exists in the hidden target object. By scanning the angle of incidence of the illuminating laser beam and maximizing the variation of the detected fluorescence signal from the object, as measured by a bucket detector at the front of the scattering medium, we are able to generate a tightly focused excitation spot. Thereafter, an image is obtained by scanning the focus over the object. The requirements for applicability of the method are discussed.**

Inhomogeneous samples randomly scatter and diffuse light, preventing the formation of diffraction-limited spots within them and blurring images. Even a very thin layer of a scattering material can appear opaque and mask objects behind it. Significant effort has been devoted in recent years to develop methods such as wavefront shaping and ghost imaging so as to enable imaging of objects behind or within random scattering media [1-5]. However, most of these methods are invasive since they require the presence of a detector [6-7] or some other form of reporter (e.g. "guide star") [8] to be placed behind or within the scattering layer. In 2012, Bertolloti et al. [9] reported an optical method that allows non-invasive imaging of a fluorescent object that is completely hidden behind a thin scattering layer. They scanned the angle of the illuminating laser beam and collected the total fluorescence emitted from the object through the scattering medium by a bucket detector. By exploiting the fact that for a thin scatterer the autocorrelation of the scattered light pattern is, to a good approximation, identical to the autocorrelation of the object's image they were able to obtain the image of the hidden object using an iterative algorithm [10], a method later extended to non-invasive imaging behind scattering media by using spatially incoherent light and a standard digital camera [11] ,and to imaging through multicore fiber bundles [12]. Speckle correlation approaches for noninvasive wide-field fluorescence imaging have also recently given in [13,14] Notably, all these methods are generally suitable for relatively simple objects due to the difficulty in inverting the autocorrelation operation, and for flat objects due to the infinite focal depth in speckle correlation based imaging. Moreover, despite the relatively low readout noise of currently available detectors, such as scientific CMOS cameras, these methods are sensitive to noise for dim objects, as the signal is spread over a large number of camera pixels. Very recently, Stern et al. [15] used a similar camera-based speckle-correlation approach to support a wavefront-shaping apparatus, allowing noninvasive, diffraction-limited focusing, which is subject to similar restrictions and limitations. In an alternative approach, Katz and coworkers have shown that focusing via wavefront shaping is possible when observation is made through a nonlinear effect such as two photon fluorescence [16]. In this case, one can monitor the total fluorescence light returning through the scattering layer, which is maximized when the excitation light is as tightly focused as possible. Using the memory effect for scanning, a full image of rather complex objects could be obtained. Moreover, once a tight focus has been formed, it can be axially scanned via wavefront shaping, enabling, in principle, imaging of three-dimensional objects. Nevertheless, the use of a nonlinear signal required the use of an ultrafast laser source and is not suitable for imaging with continuous wave laser sources. We then, similarly to the nonlinear focusing approach, scan this focus to obtain an image of the masked object.

In this work, we show that it is possible to non-invasively achieve focusing of light behind a scattering medium without the need for nonlinearity, using standard wavefront shaping techniques and a simple detection scheme using a single bucket detector. To do so, we exploit the spatial variance of the fluorescence contrast that naturally exists in most objects of interest, along with the memory effect (at least for a short range).  The concept is illustrated in Fig. 1(a). A Laser light source (outlined in green), that has passed through the scattering layer, illuminates a spatially nonuniform fluorescent object. Scanning the angle of incidence of the laser beam over a small range (of the size of several speckle grains), results in shifting of the speckle pattern due to the memory effect [17].  The total intensity of the scattered fluorescence from the object, corresponding to the convolution of the excitation pattern with the object, can be measured through the scatterer by a bucket detector, and exhibits fluctuations as we shift the speckle pattern. If the speckle pattern is relatively uniform over the object, we expect only small changes in the measured light intensity vs. angle (see Fig 1(b)); however, a focused light beam will produce large variations (Fig 1(c)). For maximizing the amplitude of the variations, the input field illuminating the random medium is controlled by a two-dimensional spatial light modulator (SLM). By using an iterative optimization algorithm [GA, for more details see Refs. 18-19], the spatial phase pattern applied by the SLM is optimized for increasing the standard deviation of the measured light intensity as a function of the angle of scanning (Ө). As the optimization process proceeds, a tight bright spot is being formed, which can then be raster-scanned (and, in the case of 3D objects, axially shifted) to generate an image (for a detailed experiment setup see Supplement 1).

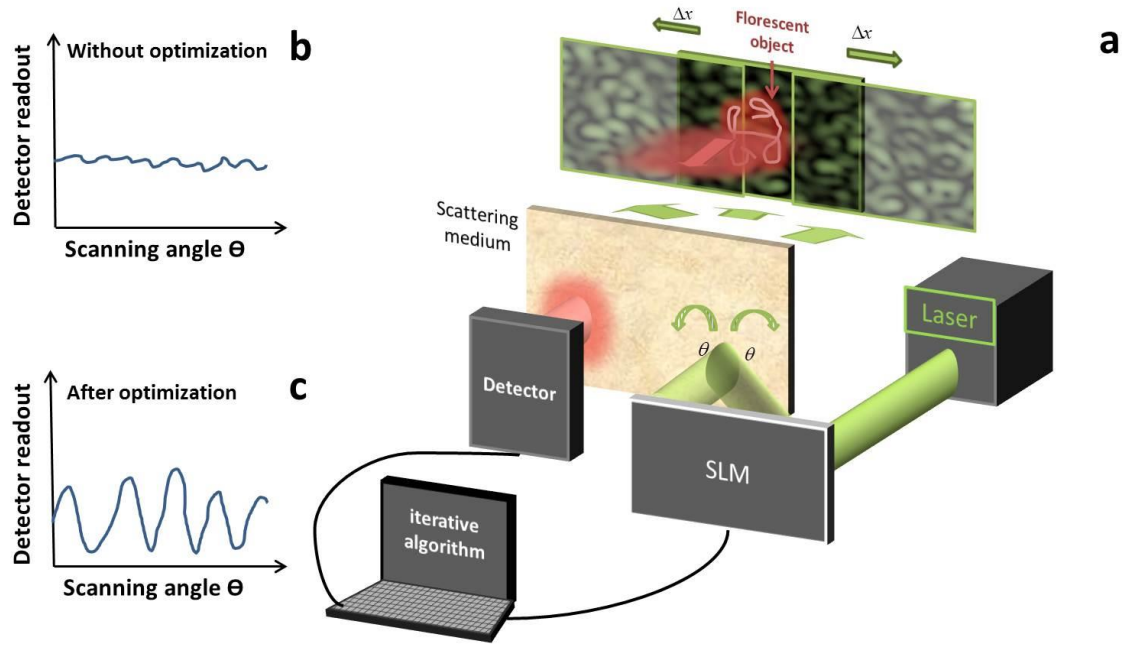

Fig. 1. (a) Experiment set up (for the detailed setup see Supplement1). A laser light illuminates a thin scatterer (ground-glass) medium at an angle ϴ. A hidden object behind the scattering medium is excited by the speckle field and fluoresces. The scattered fluorescence light is measured through the scatterer at the front with a bulk detector. By scanning the illuminating angle ϴ, the speckle pattern is shifted over the object. The total intensity of the scattered fluorescence light will show small fluctuations as we shift the speckle pattern (b). For maximizing the amplitude of the variations in (b), the input field illuminating the random medium is controlled by a two-dimensional spatial light modulator (SLM) that is imaged on the diffuser, and an iterative algorithm finds the spatial phase pattern that increases the standard deviation of the measured scattered light intensity (c). (The plots in (b) and (c) are illustrations and are for demonstration only).

To demonstrate this concept we construct a complex 2D nonuniform fluorescent layer, as seen in Fig 2(a). It contains 10 ㎛ fluorescent beads (Micro particles based on melamine resin, stained with rhodamine B) that are randomly dispersed on a glass slide. Notably, the characteristic feature size on the glass slide is larger than the bead diameter due to aggregation. The beads are excited with a continuous wave 532 nm laser and fluorescence is emitted around 590 nm. We use low numerical aperture for collecting the fluoresce light by placing the object at the far field, 8 cm away from the diffuser. In this manner we ensure that the characteristic feature size (which limits the resolution of our method) is smaller than a single speckle grain. We then scan the speckles by an angular range that is equivalent to a size of five speckle grains for introducing the spatial fluorescent fluctuations on which the genetic algorithm works. By scanning the speckle pattern (Fig 2(b)) over this object and maximizing the variation of the fluorescent signal during the scan, a clearly visible enhanced diffraction limited spot is formed behind the scattering medium (Fig 2 (c)). The images presented in Fig. 2 are taken in transmission mode (so as to obtain a clear image) but are only for verification of the technique and are not used in the optimization process.

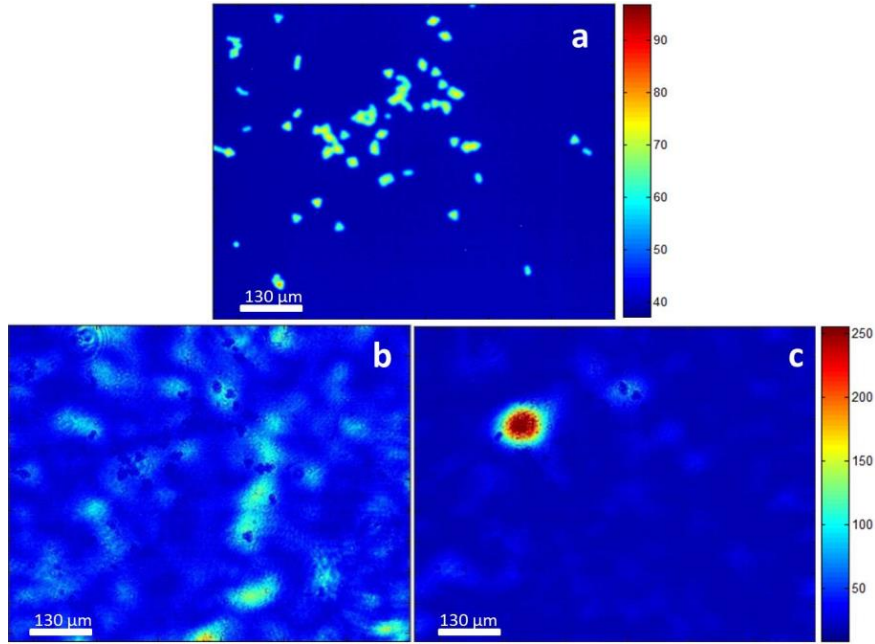

Fig. 2. (a) Fluorescent hidden object; the image of the sample is taken from the transmission side of the diffuser, after averaging over 50 speckles illuminations (resulting in a constant excitation density). (b) Transmitted speckle pattern that illuminates the sample before the optimization. Due to absorption of the laser by the object, a dark print of the object is visible, which makes it convenient to visualize the proportions between the speckle field and the features of the object. (c) A bright spot as seen from the transmitted side of the diffuser (behind the scattering medium), after maximizing the variance of the fluorescent signal from the hidden object.

Next, we use the focused excitation spot to obtain the image. For reconstructing the image, we raster scan the bright spot over the object and map the detected intensity for every position. To improve the contrast, we repeat the raster scan without the optimized bright spot (when the speckle field is much more uniform over the object) and generate a background image which is then subtracted from the retrieved mapping. To compare the obtained image with a "ground truth", we measure a diffraction limited image of the object from behind the scattering medium at a spatial resolution much higher than that achievable through the scattering layer (Fig 2(a)) and convolve it with the diffraction-limited excitation PSF shown in Fig (2(c)). The results are represented in Fig 3, where Fig. 3(a) presents the retrieved image and Fig. 3(b) presents the PSF convolved with the ground truth image. As can be seen, the noninvasive retrieved image clearly resembles the diffracted- limited object with high accuracy, but also with some added noise to the background speckles following optimization.

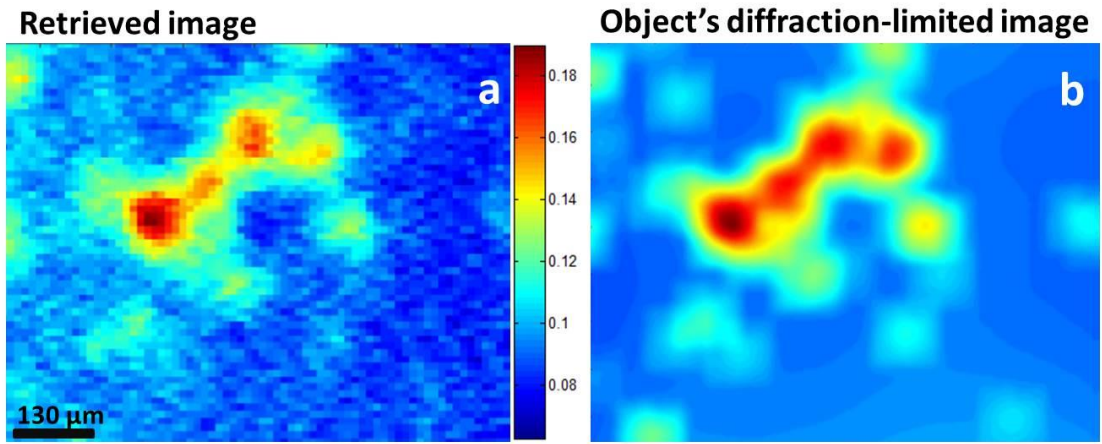

Fig. 1. (a) Reconstructed image of the hidden object as was retrieved by scanning the focus point over the sample, and collecting the fluorescent at the front side of the diffuser noninvasively. (b) Diffraction limited image of the hidden object taken from the transmitted side of the diffuser.

This demonstration shows the feasibility of our suggested noninvasive imaging method to visualize two-dimensional high contrast objects by means of wavefront shaping. Generating a tight bright spot is a direct result of the fact that the variance of the convolution of sample with the excitation intensity point spread function is maximized for a delta-function excitation. The resolution of the retrieved image is inherently determined by the illuminating speckle size over the object. For smaller speckle grain size, a finer detailed image will be obtained; however it will also require a higher number of controllers of the SLM to reach the same contrast between the bright spot and the background [20]. Additionally, the field of view of the image is determined by the memory effect range (although we note that within that range the excitation spot could potentially be adaptively corrected locally, further extending the relevant imaging range). Since the optimization method presented above inherently relies on the utilization of the spatial inhomogeneity of the object fluorescence, it is essential to consider the effect of the object geometry on the success of the method. Clearly, in case of a completely homogenous sample, this method fails, since the convolution of any speckle pattern with the object would yield an identical signal. On the other hand, a very sparse object (composed, for example, from a few fluorescent spots which between them, the distance is significantly longer than the distance at which the speckles are shifted during the scan) will also likely not yield a single bright spot. In this case, the optimization process can potentially converge to a solution which contains several bright foci – each one of them is located on a different fluorescent spot of the object. In order to verify this, we performed a computer simulation for studying the efficacy of our method on different object's geometries. The object is represented by a random binary matrix of zeros and ones (where the ones elements represent the fluorescent beads), and its geometry is controlled by a degree of sparsity that we set. Fig 4 (a) (c) and (e) are representative geometries we study, with low (66% of the element are zeros), high (95% are zeros )and very high degree (99.7% are zeros) of sparsity respectively. We convolve the speckle pattern with the object matrix to simulate the fluorescent signal (see simulation details on Supplement 1). We then, calculate the optimal setting of the SLM phases that will maximize the signal variations as we shift the speckles over the object using the same optimization algorithm used in the experiments.

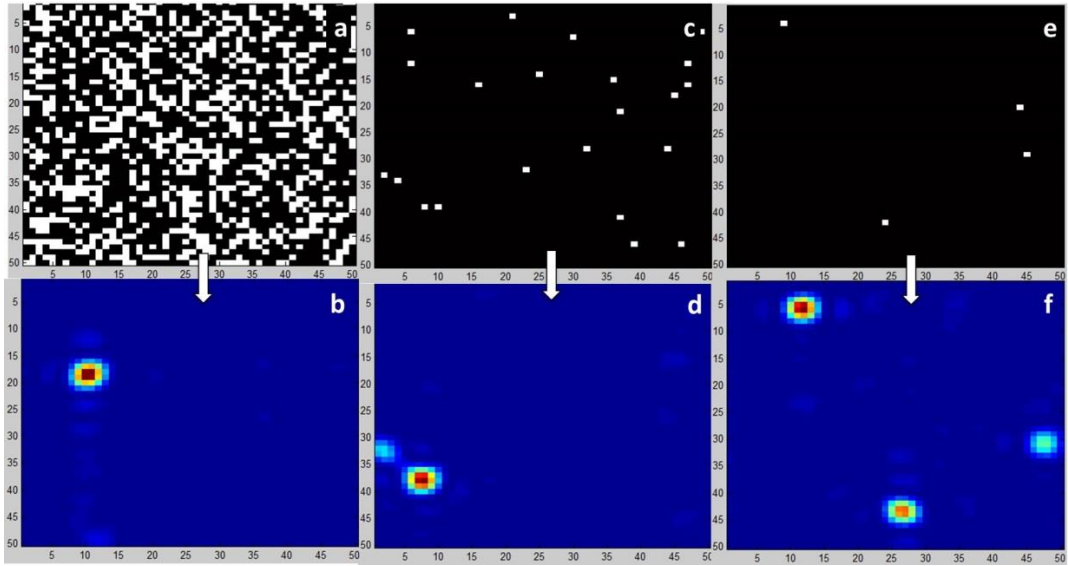

Fig. 4. Simulation results of optimization over different object's geometries. (a),(c) and (e) Low sparsity, high sparsity and very high sparsity objects respectively are represented as binary matrix where white represent fluorescence spots and black is dark spots. For low and high sparsity the optimization results with a single focal point (b) and (d) . With very sparse object the optimization results with at least two enhanced focal spots (f). The foci are located at the same position where bright spots appear on the object.

Results of the simulation are depicted in Fig. 4. when the sample is random with degree of sparsity lower than 95% (as in the case experimentally tested in Fig. 2), there is one focal point (Fig 4 (b)). Note that in this case the position of the bright spot is difficult to predict a-priori. However, when the sample is highly sparse, other focal points can be seen (Fig. 4 (f)). It is also noticeable that these focal points are positioned on bright areas of the sample, such that the optimized pattern is a partial image of the original object. In case where the object is sparse but not highly sparse(as in Fig4(c)), most of algorithm realizations yield one single bright spot . Higher degree of sparsity (reaching up to 99.5% of zeros) also converges to one single point in most of the cases .

In summary, we have demonstrated focusing and imaging through a thin scattering medium noninvasively by using wavefront shaping technique that utilizes the intrinsic variability of the fluorescent sample. Our method is applicable with linear excitation and does not require the use of an imaging detector, since no spatial information is used in generating feedback signal. Bucket detectors, as used in our scheme, are highly advantageous relative to imaging detectors in terms of signal to noise, allowing the retrieval of dim objects or when the collection efficiency of the emitted fluorescence is low (as is the case here). We have also shown that as long as the sparsity of the object is sufficiently small, such that the characteristic distance between fluorescent regions is smaller than the scanning range, we were able to generate a single bright focal spot and retrieve an image of relatively complex object. This method is in principle applicable to 3D objects, although the focal plane at which the focal spot will be generated is a-priori unknown. On the other hand, our method is still limited by the memory effect range which is determined by the thickness and roughness of the scattering media. This will limit the size of the object that can be scanned.

The typical time scale of such an optimization experiment is typically tens of minutes, predominantly due to the relatively low refresh rate of the liquid-crystal SLM. Notably, these time scales can be reduced by using faster SLMs such as ones based on digital micromirror devices and potentially also by better optimization algorithms which were not examined in the frame of our work. This method may also be able to yield useful information about the statistical properties of the fluorophore distribution in a scattering sample even without the need for imaging.

See Supplement 1 for supporting content.

## Supplement 1

***Experiment Apparatus***. Figure S1 illustrates in details the demonstrative experimental set up as described in the main text. Since we are measuring weak scattered signal, we are using lock-in amplifier for better signal readout. The light source is 532nm laser. The beam passes through chopper (this is the reference signal of the lock-in amplifier) and is then expanded and reflected off the SLM (HAMAMATSU, LCOS-SLM X10468), which is then imaged on the diffuser. The SLM resolution is 600X800 pixels. The SLM is divided into 60X80 effective segments; each block is controlled independently to apply a phase between 0 to $2\pi$. The reflected beam is passing through a telescope that de-magnifies the SLM surface by a factor 8 on the diffuser (LSD, angle: 5°). Light scattered by the diffuser creates a speckle pattern that excites a florescent object. The speckle field is shifted over the object by controlling two galvo mirrors that are imaged on the SLM plane. The florescent signal passes through the diffuser and reflected off dichroic mirror into a photomultiplier detector (PMT). Optimization of the phase pattern is performed with a genetic algorithm with the target of maximizing the variance of the florescent signal as the speckle pattern is shifted.

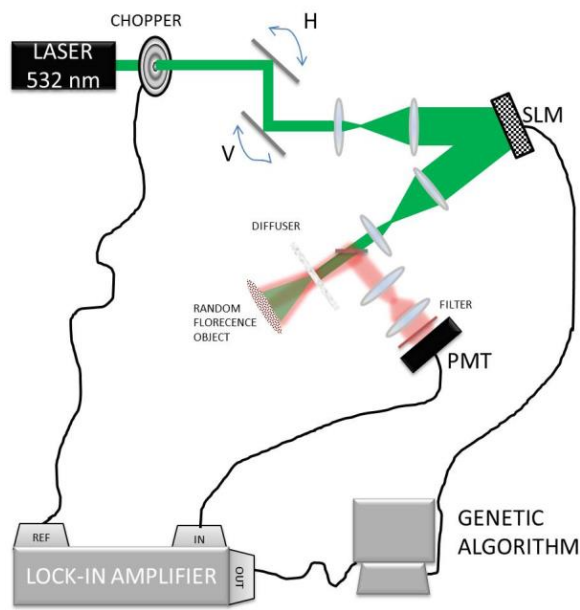

Fig. S1. Experiment apparatus

***Simulation***. The speckle pattern was simulated by applying 2D Fourier transform to a random phase matrix of size 256X256. The pattern at the far field is multiplied by random binary matrix of zeros and ones with variant sparsity that simulates the florescent object. The speckle pattern is shifted over the binary matrix. The detected signal is the sum of all elements in the multiplication matrix of the object by the speckles. Genetic algorithm (Matlab toolbox) finds the optimal setting of the SLM phases that will maximize the signal variations as we shift the speckles over the object.